\documentclass[english,showpacs,preprint,superscriptaddress]{revtex4-1}
\usepackage{lmodern}
\usepackage{lmodern}
\usepackage[T1]{fontenc}
\usepackage{bm}
\usepackage{amsmath}
\usepackage{amssymb}
\usepackage{graphicx}
\usepackage{esint}

\makeatletter

\providecommand{\tabularnewline}{\\}

\usepackage{amsthm}\usepackage{latexsym}\usepackage{bm}\usepackage{amsfonts}

\usepackage{babel}

\usepackage{babel}
\usepackage{hyperref}

\makeatother

\usepackage{babel}
\begin{document}

\title{Lorentz Covariant Canonical Symplectic Algorithms for Dynamics of
Charged Particles}

\author{Yulei Wang}

\affiliation{School of Nuclear Science and Technology and Department of Modern
Physics, University of Science and Technology of China, Hefei, Anhui
230026, China}

\affiliation{Key Laboratory of Geospace Environment, CAS, Hefei, Anhui 230026,
China}

\author{Jian Liu}

\email{corresponding author: jliuphy@ustc.edu.cn}

\affiliation{School of Nuclear Science and Technology and Department of Modern
Physics, University of Science and Technology of China, Hefei, Anhui
230026, China}

\affiliation{Key Laboratory of Geospace Environment, CAS, Hefei, Anhui 230026,
China}

\author{Hong Qin}

\affiliation{School of Nuclear Science and Technology and Department of Modern
Physics, University of Science and Technology of China, Hefei, Anhui
230026, China}

\affiliation{Plasma Physics Laboratory, Princeton University, Princeton, NJ 08543,
USA}
\begin{abstract}
In this paper, the Lorentz covariance of algorithms is introduced.
Under Lorentz transformation, both the form and performance of a Lorentz
covariant algorithm are invariant. To acquire the advantages of symplectic
algorithms and Lorentz covariance, a general procedure for constructing
Lorentz covariant canonical symplectic algorithms (LCCSA) is provided,
based on which an explicit LCCSA for dynamics of relativistic charged
particles is built. LCCSA possesses Lorentz invariance as well as
long-term numerical accuracy and stability, due to the preservation
of discrete symplectic structure and Lorentz symmetry of the system.
For situations with time-dependent electromagnetic fields, which is
difficult to handle in traditional construction procedures of symplectic
algorithms, LCCSA provides a perfect explicit canonical symplectic
solution by implementing the discretization in 4-spacetime. We also
show that LCCSA has built-in energy-based adaptive time steps, which
can optimize the computation performance when the Lorentz factor varies.
\end{abstract}
\maketitle

\section{Introduction\label{sec:Introduction}}

The advanced structure-preserving geometric algorithms have stepped
into the field of plasma physics and attracted more and more attentions
in recent years \cite{Qin_VariatianalSymlectic_2008,Qin_VSSymp_in_Tokamak,LiJinXing_GC_Symp_2011,Guan_Qin_Sympletic_RE,LiuJian_RE_Positron_2014,Jianyuan_Multi_sympectic_2013,XiaoJY_PIC_wave_2015,ExpNoncanonicalS_2015,CSPIC_2016,CollisionlessScater_NF_Letter_2016,RELong_WangYulei2016}.
Through preserving different geometric structures, such as the phase-space
volume, the symplectic structure, and the Poisson structure, geometric
algorithms possess long-term numerical accuracy and stability and
have shown powerful capabilities in dealing with multi-scale and nonlinear
problems. Volume-preserving algorithms (VPA) of different orders for
both non-relativistic and relativistic full-orbit dynamics of charged
particles have been constructed in several publications \cite{HeYang_Spliting_2015,Ruili_VPA_2015,HeYang_HigherOrderVPA_2016}.
A Poisson-preserving algorithm for solving the Vlasov-Maxwell system
is built through splitting the Hamiltonian and using the Morrison-Marsden-Weistein
bracket \cite{HeYang_HamiltonTimeInt_VMs_2015}. As an important aspect
of geometric algorithms, symplectic methods have produced fruitful
results. For gyro-center dynamics of charged particles, variational
symplectic methods have been studied and applied to plasma simulations
\cite{Qin_VariatianalSymlectic_2008,Qin_VSSymp_in_Tokamak,LiJinXing_GC_Symp_2011}.
It is also feasible to canonicalize the gyro-center equations to construct
canonical symplectic algorithms for time-independent magnetic fields
\cite{Ruili_GC_canonical_2014}. The Particle-in-Cell (PIC) method,
known as the first principle simulation method for plasma systems,
has been reconstructed by the use of different symplectic methods,
including variational symplectic method, canonical symplectic method,
and non-canonical symplectic method \cite{Jianyuan_Multi_sympectic_2013,XiaoJY_PIC_wave_2015,CSPIC_2016,ExpNoncanonicalS_2015}.
 Theoretically, symplectic methods impose numerical results with a
set of constrains, the number of which is determined by the freedom
degrees of the systems \cite{Geometric_numerical_integration,CSPIC_2016},
by preserving the global symplectic structure of the system. Correspondingly,
the global relative errors of motion constants can be restricted to
bounded small values, which enable symplectic algorithms to retain
many key properties of the origin continuous systems. However, another
essential geometric property of physical systems has long been ignored
in structure-preserving algorithms, i.e., the Lorentz covariance.
The lack of Lorentz covariance leads to inconsistent numerical solutions
in different inertial frames. In this paper, we equip the symplectic
algorithm with the Lorentz covariance to obtain better performances.

As an intrinsic property of continuous physical systems, the Lorentz
covariance has become a common sense in modern physics, which states
that the physical rules and events keep invariant under Lorentz transformation
\cite{Jackson_electrodynamics}. It is also important for algorithms
to satisfy the Lorentz covariance. Similar to continuous covariant
system, Lorentz covariant algorithms have invariant forms and describe
invariant processes under Lorentz transformation. The Lorentz invariance
of each one-step map ensures the reference-independence of numerical
results, which leads to that the numerical properties, such as stability,
convergence, and consistency are also independent with the choice
of reference frames. In applications, the Lorentz covariant algorithms
make it convenient and safe to adopt the same set of discretized equations
in all inertial frames.

The combination of Lorentz covariance and symplectic method can generate
algorithms possessing benefits from the both. If the long-term numerical
accuracy and stability are unavailable, Lorentz covariant algorithms
cannot guarantee the long-term correctness of simulations, even though
the results are reference-independent. On contrary, although symplectic
methods without Lorentz covariance have long-term conservativeness
and stability, they break the Lorentz symmetry of the original continuous
systems and produce inconsistent numerical solutions in different
inertial frames. On the other hand, it is difficult to construct conventional
symplectic algorithms for time-dependent Hamiltonian systems. Meanwhile,
it is not straightforward to develop conventional symplectic algorithms
with optimized adaptive time steps. These two problems can be solved
automatically by the construction of Lorentz covariant symplectic
algorithms. Covariant algorithms directly iterate geometric objects
in 4-spacetime and discretize the worldlines with respect to the discrete
proper time $\tau$. Consequently, the time, $t$, as a component
of the 4-spacetime, plays the same role as spatial coordinates in
time-dependent Hamiltonians. Taking the place of $t$, the proper
time is employed as the dynamical parameter and leads to proper-time-independent
Hamiltonians for time-dependent systems. Because covariant algorithms
directly discretize the worldline, one can obtain energy-based adaptive-time-step
symplectic schemes given the fixed proper-time step $\Delta\tau=\Delta t/\gamma$.
The adaptive time step can improve the performance of symplectic algorithms
when the Lorentz factor varies.

To endue symplectic algorithms with Lorentz covariance, a straightforward
way is to start from the view point of geometry. The Lorentz covariant
systems reside in the 4-dimentional spacetime. Considering the reference-independence,
the Lorentz covariant discretized equations can be regarded as the
one-step maps of geometric objects in spacetime. As a result, if one
starts from covariant continuous geometric equations, and discretizes
these equations without breaking the integrity of all the geometric
objects, the Lorentz covariance can be naturally inherited. The canonical
symplectic methods directly deal with the Hamiltonian equations of
physical systems. During discretization, the symplectic structure
of Hamiltonian equations is retained, and each of the physical quantities
is treated as an inseparable discretized geometric object, updated
at different proper-time steps \cite{Geometric_numerical_integration}.
It is readily to see that the canonical symplectic method provides
a convenient way to combine the symplectic method and the Lorentz
covariance. Here we summarize a general procedure for constructing
Lorentz covariant canonical symplectic algorithms (LCCSA), namely,
1) to write down the covariant geometric Hamiltonian equation for
a target physical system in 4-dimentional spacetime, 2) to discretize
the Hamiltonian equations by using a canonical symplectic scheme,
such as Euler-symplectic scheme and implicit mid-point symplectic
scheme, described by geometric objects in 4-spacetime.

Following this procedure, we construct an explicit LCCSA for the simulation
of relativistic dynamics of charged particles. Compared with a non-covariant
algorithm, LCCSA exhibits the reference-independent form and good
long-term performances in different Lorentz frames. As a symplectic
algorithm, LCCSA shows outstanding long-term numerical accuracy than
a covariant fourth-order Runge-Kutta algorithm (RK4). Meanwhile, LCCSA
can automatically adjust the time step-length according to the energy
of a particle and guarantee the approximate constant time-sampling
number in one gyro-period. The performance in simulating energy-changing
processes can be improved. As examples, both the computation efficiency
for simulating acceleration and braking processes of charged particle
by use of LCCSA are optimized compared with those fixed-time-step
algorithms. 

The rest part of this paper is organized as follows. The definition
and properties of Lorentz covariant symplectic algorithms are introduced
in Sec.\,\ref{sec:Lorentz-Covariant-Algorithms}. The detailed procedure
of constructing an explicit LCCSA is explained in Sec.\,\ref{sec:Construction-CCSA}.
In Sec.\,\ref{sec:Drawbacks-of-Non-covariant}, the performances
of LCCSA are exhibited through several typical numerical cases. We
summarize this article in Sec.\,\ref{sec:Conclusions}.

\section{Lorentz Covariant Symplectic Algorithms\label{sec:Lorentz-Covariant-Algorithms}}

Before introducing Lorentz covariant symplectic algorithms, we first
provide the rigorous definition of Lorentz covariant algorithm. For
a given continuous Lorentz covariant system $\mathbf{F}$, an algorithm
$A$ is called Lorentz covariant if and only if it satisfies 
\begin{equation}
\mathcal{D}_{A}\circ\mathcal{T}_{L}\mathbf{F}=\mathcal{T}_{L}\circ\mathcal{D}_{A}\mathbf{F}\,,\label{eq:CovAlgorithmDef}
\end{equation}
where $\mathcal{T}_{L}$ denotes the Lorentz transformation operator,
$\mathcal{D}_{A}$ denotes the discretization operator determined
by the algorithm $A$, the operation ``$\circ$'' means composite
mapping, and $L$ denotes the Lorentz transformation matrix satisfying
$L^{T}gL=g$, where $g$ is the Lorentz metric tensor of the 4-dimentional
spacetime \cite{Jackson_electrodynamics,QinHong_Intro_GeoGyroKinetics,QinHong_Edge_Plasma_GeoGyroKinetics}.
Generally speaking, $L$ can be both proper Lorentz transformation
($\mathrm{det}\,L=+1$) and improper Lorentz transformation ($\mathrm{det}\,L=-1$).
A Lorentz transformation includes the rotation and the Lorentz boost
of inertial frames \cite{Jackson_electrodynamics}. Suppose that $A$
is applied to system $\mathbf{F}$ in the inertial frame $\mathcal{O}$,
the first operation on the right-hand side of Eq.\,\ref{eq:CovAlgorithmDef},
$\phi_{A}=\mathcal{D}_{A}\mathbf{F}$, gives a realization of algorithm
$A$, i.e., a set of discrete equations in this frame. In another
inertial frame $\mathcal{O}'$ moving with speed $\bm{\beta}$ relative
to $\mathcal{O}$, this discrete system are described by discrete
equations $\phi_{A}'=\mathcal{T}_{L}\phi_{A}$ following the Lorentz
transformation of $\phi_{A}$. On the left-hand side of Eq.\,\ref{eq:CovAlgorithmDef},
because the original system $\mathbf{F}$ is Lorentz covariant, $\mathbf{F}'=\mathcal{T}_{L}\mathbf{F}$
takes the same form as $\mathbf{F}$. Consequently, the realization
of algorithm $A$ on $\mathbf{F}'$, i.e., $\xi_{A}=\mathcal{D}_{A}\mathbf{F}'$,
also takes the same form as $\phi_{A}$ except that the physical quantities
in $\xi_{A}$ are observed in the frame $\mathcal{O}'$. So Eq.\,\ref{eq:CovAlgorithmDef}
concludes that the discrete equations generated by a Lorentz covariant
algorithm $A$ have the invariant form and provides the same discretized
system in different Lorentz inertial frames. 

To make the picture of covariant algorithms clearer, for comparison,
we investigate an example of a non-covariant algorithm, i.e., the
VPA for relativistic charged particles dynamics as constructed in
\cite{Ruili_VPA_2015}. This algorithm has been applied to the study
of long-term dynamics of runaway electrons in tokamaks and shown its
outstanding long-term numerical accuracy \cite{CollisionlessScater_NF_Letter_2016,RELong_WangYulei2016}.
However, its non-Lorentz-covariant property can be proved according
to Eq.\,\ref{eq:CovAlgorithmDef} as follows. The target continuous
system is the relativistic Lorentz force equations $\mathbf{F}_{L3}$
\begin{equation}
\frac{\mathrm{d}\mathbf{x}}{\mathrm{d}t}=\frac{\mathbf{p}}{\gamma}\,,\label{eq:Lorentz3Dx}
\end{equation}
\begin{equation}
\frac{\mathrm{d}\mathbf{p}}{\mathrm{d}t}=\mathbf{E}+\frac{\mathbf{p}\times\mathbf{B}}{\gamma}\,,\label{eq:Lorentz3Dp}
\end{equation}
where $\mathbf{x}$ is the position, $\mathbf{p}$ is the mechanical
momentum, $\gamma=\sqrt{1+p^{2}}$ is the Lorentz factor, and $\mathbf{E}$
and $\mathbf{B}$ are respectively electric and magnetic fields. Notice
that all the physical quantities in this paper are normalized according
to Tab.\,\ref{tab:Units} unless noted otherwise. As a common wisdom,
$\mathbf{F}_{L3}$ is Lorentz covariant. We will show that $\mathcal{D}_{VPA}\circ\mathcal{T}_{L}\mathbf{F}_{L3}\neq\mathcal{T}_{L}\circ\mathcal{D}_{VPA}\mathbf{F}_{L3}$.

\begin{table}
\begin{tabular}{|c|c|c|}
\hline 
Names & Symbols & Units\tabularnewline
\hline 
\hline 
Time, Proper Time, Gyro-period & $t$, $\tau$, $T_{ce}$ & $\mathrm{m_{0}}/\mathrm{e}B_{0}$\tabularnewline
\hline 
Position & $\mathbf{x}$ & $\mathrm{m_{0}c}/\mathrm{e}B_{0}$\tabularnewline
\hline 
Mechanical/Canonical Momentum & $\mathbf{p}$, $\mathbf{P}$ & $\mathrm{m_{0}c}$\tabularnewline
\hline 
Velocity & $\mathbf{v}$, $\bm{\beta}$ & $\mathrm{c}$\tabularnewline
\hline 
Electric field & $\mathbf{E}$ & $B_{0}\mathrm{c}$\tabularnewline
\hline 
Magnetic field & $\mathbf{B}$ & $B_{0}$\tabularnewline
\hline 
Vecter field & $\mathbf{A}$ & $\mathrm{e}/\mathrm{m}_{0}\mathrm{c}$\tabularnewline
\hline 
Scalar field & $\phi$ & $\mathrm{e}/\mathrm{m}_{0}\mathrm{c}^{2}$\tabularnewline
\hline 
Hamiltonian & $\mathcal{H}$ & $\mathrm{m_{0}c^{2}}$\tabularnewline
\hline 
\end{tabular}

\caption{Units of all the physical quantities used in this paper. $\mathrm{m}_{0}$
is the rest mass of a particle, $\mathrm{e}$ is the elementary charge,
$\mathrm{c}$ is the speed of light, and $B_{0}$ is the given reference
magnetic field.\label{tab:Units}}
\end{table}

Firstly, we derive the discrete system $\phi'_{VPA}=\mathcal{T}_{L}\circ\mathcal{D}_{VPA}\mathbf{F}_{L3}$.
In the reference frame $\mathcal{O}$, by applying the discrete operator
to $\mathbf{F}_{L3}$ we obtain $\phi_{VPA}=\mathcal{D}_{VPA}\mathbf{F}_{L3}$,
which is a set of difference equations in $\mathcal{O}$ and can be
written explicitly as \cite{Ruili_VPA_2015},
\begin{equation}
t_{k+1}=t_{k}+\Delta t\,,\label{eq:VPAt}
\end{equation}
\begin{equation}
\mathbf{x}{}_{k+1}=\mathbf{x}{}_{k}+\Delta t\frac{\mathbf{p}{}_{k}}{\gamma{}_{k}}\,,\label{eq:VPAp_x}
\end{equation}
\begin{equation}
\mathbf{p}{}_{k+1}=\mathbf{p}{}_{k}+\mathbf{W}\,,\label{eq:VPAp}
\end{equation}
 where $\gamma_{k}=\sqrt{1+p_{k}^{2}}$, and $\mathbf{W}\left(t_{k+1},\mathbf{x}_{k+1},\mathbf{p}_{k},\mathbf{E}_{k+1},\mathbf{B}_{k+1},\Delta t\right)$
is a function given by
\begin{equation}
\mathbf{W}=\Delta t\mathbf{E}_{k+1}+\left(D\hat{\mathbf{B}}_{k+1}+dD\hat{\mathbf{B}}_{k+1}^{2}\right)\left(\mathbf{p}_{k}+\frac{\Delta t}{2}\mathbf{E}_{k+1}\right)\,,\label{eq:VPA_W}
\end{equation}
 where $\mathbf{E}_{k+1}=\mathbf{E}\left(t_{k+1},\mathbf{x}_{k+1}\right)$,
$\mathbf{B}_{k+1}=\mathbf{B}\left(t_{k+1},\mathbf{x}_{k+1}\right)$,
$d=\Delta t/\left[2\sqrt{1+\left(\mathbf{p}_{k}+\Delta t\mathbf{E}_{k+1}/2\right)^{2}}\right]$,
and $D=2d/\left(1+d^{2}B_{k+1}^{2}\right)$, and in Cartesian coordinate
system $\hat{\mathbf{B}}$ is defined as
\begin{equation}
\hat{\mathbf{B}}=\left(\begin{array}{ccc}
0 & B_{z} & -B_{y}\\
-B_{z} & 0 & B_{x}\\
B_{y} & -B_{x} & 0
\end{array}\right)\,.\label{eq:Bhat}
\end{equation}
Then, we transform $\phi_{VPA}$ into another frame $\mathcal{O}'$.
We suppose that $\mathcal{O}'$ moves with a fixed speed $\bm{\beta}=\left(\beta_{1},\beta_{2},\beta_{3}\right)$
relative to $\mathcal{O}$. In this case, the Lorentz matrix $L$
denotes the Lorentz boost matrix which can be written explicitly in
the Cartesian coordinate system as, 
\begin{equation}
L=\left(\begin{array}{cccc}
\Gamma & -\Gamma\beta_{1} & -\Gamma\beta_{2} & -\Gamma\beta_{3}\\
-\Gamma\beta_{1} & 1+\frac{\left(\Gamma-1\right)\beta_{1}^{2}}{\beta^{2}} & \frac{\left(\Gamma-1\right)\beta_{1}\beta_{2}}{\beta^{2}} & \frac{\left(\Gamma-1\right)\beta_{1}\beta_{3}}{\beta^{2}}\\
-\Gamma\beta_{2} & \frac{\left(\Gamma-1\right)\beta_{1}\beta_{2}}{\beta^{2}} & 1+\frac{\left(\Gamma-1\right)\beta_{2}^{2}}{\beta^{2}} & \frac{\left(\Gamma-1\right)\beta_{2}\beta_{3}}{\beta^{2}}\\
-\Gamma\beta_{3} & \frac{\left(\Gamma-1\right)\beta_{1}\beta_{3}}{\beta^{2}} & \frac{\left(\Gamma-1\right)\beta_{2}\beta_{3}}{\beta^{2}} & 1+\frac{\left(\Gamma-1\right)\beta_{3}^{2}}{\beta^{2}}
\end{array}\right)\,,\label{eq:Aboost-1}
\end{equation}
where $\beta=\left|\bm{\beta}\right|$, and $\Gamma=1/\sqrt{1-\beta^{2}}$
is the Lorentz factor of frame $\mathcal{O}'$. Without loss of generality,
we set $\bm{\beta}$ as $\left(\beta,0,0\right)$. Substituting $\mathbf{x}_{k}$,
$\mathbf{p}_{k}$, $\gamma_{k}$, $\Delta t$, $\mathbf{E}_{k}$ and
$\mathbf{B}_{k}$ in Eq.\,\ref{eq:VPAp_x} and Eq.\,\ref{eq:VPAp}
by
\begin{equation}
t_{k}=\Gamma\left(t'_{k}+\beta x'_{k}\right)\,,\label{eq:Lt}
\end{equation}
\begin{equation}
x_{k}=\Gamma\left(\beta t'_{k}+x'_{k}\right)\,,\label{eq:Lx}
\end{equation}
\begin{equation}
y_{k}=y_{k}'\,,\label{eq:Ly}
\end{equation}
\begin{equation}
z_{k}=z'_{k}\,,\label{eq:Lz}
\end{equation}
\begin{equation}
\gamma_{k}=\Gamma\left(\gamma'_{k}+\beta p_{x,k}'\right)\,,\label{eq:Lgamma}
\end{equation}
\begin{equation}
p_{x,k}=\Gamma\left(\beta\gamma'_{k}+p'_{x,k}\right)\,,\label{eq:Lpx}
\end{equation}
\begin{equation}
p_{y,k}=p'_{y,k}\,,\label{eq:Lpy}
\end{equation}
\begin{equation}
p_{z,k}=p'_{z,k}\,,\label{eq:Lpz}
\end{equation}
\begin{equation}
\Delta t=\frac{\gamma_{k}}{\gamma'_{k}}\Delta t'=\frac{\Gamma\left(\gamma'_{k}+\beta p_{x,k}'\right)}{\gamma'_{k}}\Delta t'\,,\label{eq:Ldt}
\end{equation}
\begin{equation}
\mathbf{E}_{k}=f_{E}\left(\mathbf{E}_{k}',\mathbf{B}_{k}'\right)\,,\label{eq:LE}
\end{equation}
\begin{equation}
\mathbf{B}_{k}=f_{B}\left(\mathbf{E}_{k}',\mathbf{B}_{k}'\right)\,,\label{eq:LB}
\end{equation}
where $f_{E}$ and $f_{B}$ are the Lorentz transformation functions
for electric and magnetic fields \cite{Jackson_electrodynamics}.
After simplification, the difference equations in $\mathcal{O}'$,
$\phi'_{VPA}=\mathcal{T}_{L}\phi_{VPA}$, becomes
\begin{equation}
t'_{k+1}=t'_{k}+\Delta t'\,,\label{eq:phit}
\end{equation}
\begin{equation}
\mathbf{x}'{}_{k+1}=\mathbf{x}'{}_{k}+\Delta t'\frac{\mathbf{p}'{}_{k}}{\gamma'{}_{k}}\,,\label{eq:phip_x}
\end{equation}
\begin{equation}
p'_{x,k+1}=p'_{x,k}+\beta\left(\sqrt{1+\left(p'_{k}\right)^{2}}-\sqrt{1+\left(p'_{k+1}\right)^{2}}\right)+\frac{W'_{x}}{\Gamma}\,,\label{eq:phip_px}
\end{equation}
\begin{equation}
p'_{y,k+1}=p'_{y,k}+W'_{y}\,,\label{eq:phip_py}
\end{equation}
\begin{equation}
p'_{z,k+1}=p'_{z,k}+W'_{z}\,,\label{eq:phip_pz}
\end{equation}
 where $W'_{x}$, $W'_{y}$, and $W'_{z}$ are three components of
$\mathbf{W}'\left(t'_{k+1},\mathbf{x}'_{k+1},\gamma'_{k},\mathbf{p}'_{k},\mathbf{E}'_{k+1},\mathbf{B}'_{k+1},\Delta t'\right)$
which is given by
\begin{equation}
\mathbf{W}'=\mathbf{W}\left[t_{k+1}\left(t'_{k+1},\mathbf{x}'_{k+1}\right),\mathbf{x}_{k+1}\left(t'_{k+1},\mathbf{x}'_{k+1}\right),\mathbf{p}_{k}\left(\gamma'_{k},\mathbf{p}'_{k}\right),f_{E},f_{B},\Delta t\left(\gamma'_{k},p'_{x,k},\Delta t'\right)\right]\,.\label{eq:Wp}
\end{equation}
 According to Eq.\,\ref{eq:phip_px}, $\phi'_{VPA}$ is an implicit
scheme.

Next, we derive the difference equations determined by $\xi_{VPA}=\mathcal{D}_{VPA}\circ\mathcal{T}_{L}\mathbf{F}_{L3}$.
Because Eqs.\,\ref{eq:Lorentz3Dx} and \ref{eq:Lorentz3Dp} are covariant
equations, the target continuous system in frame $\mathcal{O}'$ takes
the form $\mathbf{F}'_{L3}=\mathcal{T}_{L}\mathbf{F}_{L3}$, i.e.,
\begin{equation}
\frac{\mathrm{d}\mathbf{x}'}{\mathrm{d}t'}=\frac{\mathbf{p}'}{\gamma'}\,,\label{eq:FL3px}
\end{equation}
\begin{equation}
\frac{\mathrm{d}\mathbf{p}'}{\mathrm{d}t'}=\mathbf{E}'+\frac{\mathbf{p}'\times\mathbf{B}'}{\gamma'}\,.\label{eq:FL3pp}
\end{equation}
Discretizing $\mathbf{F}_{L3}'$ by VPA, the difference equation $\mathbf{\xi}_{VPA}=\mathcal{D}_{VPA}\mathbf{F}_{L3}'$
is given by
\begin{equation}
\mathbf{x}'{}_{k+1}=\mathbf{x}'{}_{k}+\Delta t'\frac{\mathbf{p}'{}_{k}}{\gamma'{}_{k}}\,,\label{eq:xi_x}
\end{equation}
\begin{equation}
\mathbf{p}'{}_{k+1}=\mathbf{p}'{}_{k}+\mathbf{V}'\,,\label{eq:xi_p}
\end{equation}
where $\mathbf{V}'\left(t'_{k+1},\mathbf{x}'_{k+1},\mathbf{p}'_{k},\mathbf{E}'_{k+1},\mathbf{B}'_{k+1},\Delta t'\right)=\mathbf{W}\left(t'_{k+1},\mathbf{x}'_{k+1},\mathbf{p}'_{k},\mathbf{E}'_{k+1},\mathbf{B}'_{k+1},\Delta t'\right)$.
It is obvious that $\mathbf{\xi}_{VPA}\neq\phi'_{VPA}$. That the
VPA in \cite{Ruili_VPA_2015} is not Lorentz covariant is therefore
proved. 

\begin{figure}
\includegraphics{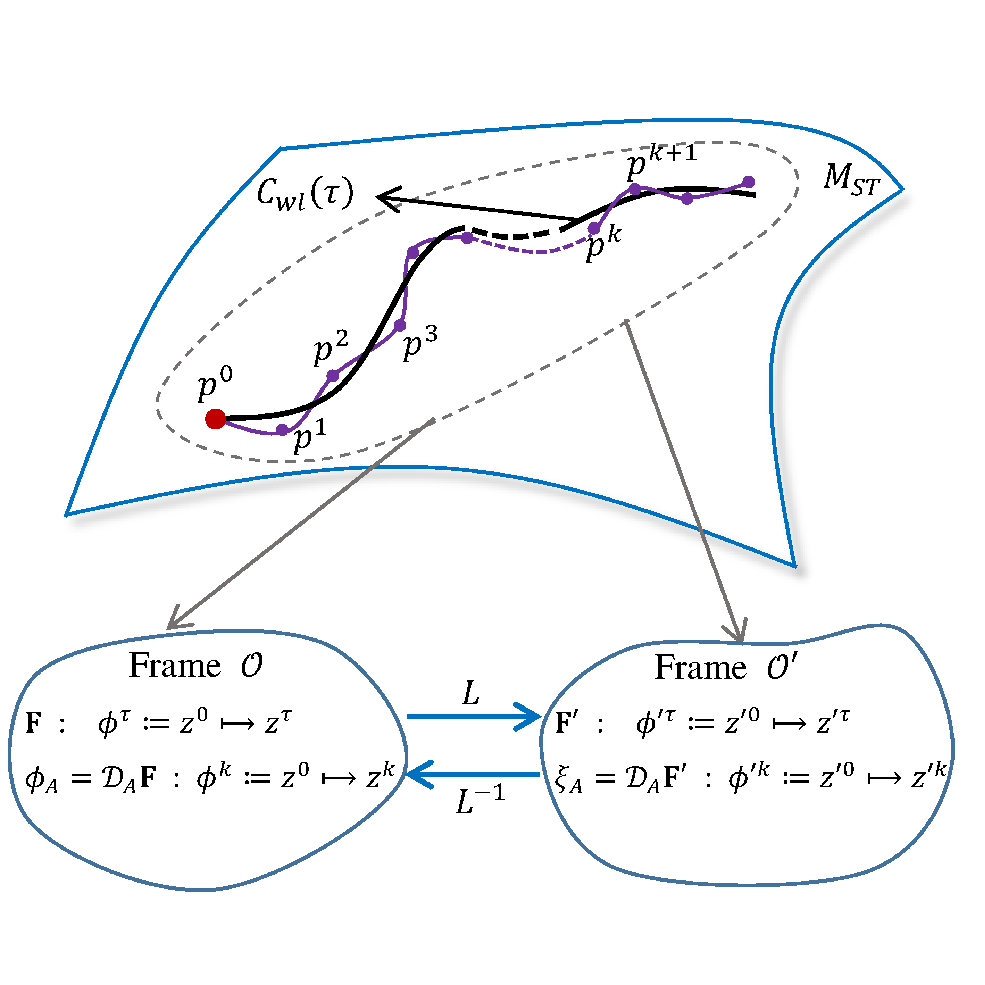}

\caption{Schematic diagram for the covariance of continuous systems and the
Lorentz covariant symplectic algorithms. $M_{ST}$ is the configuration
space of 4-spacetime. The worldline, $C_{wl}$, is denoted by the
black solid curve. The sequence of purple points, $p^{k}$, denote
the discrete approximation of $C_{wl}$. Two Lorentz frames, $\mathcal{O}$
and $\mathcal{O}'$, are chosen to express the Lorentz tranformation
relations. For covariant continuous system, $\mathbf{F}$ and $\mathbf{F}'$
have the same form, and their solutions in different reference frames
give the same worldline on $M_{ST}$ with the initial condition $p^{0}$.
Similarly, for Lorentz covariant algorithm $A$, the discrete systems
$\phi_{A}=\mathcal{D}_{A}\mathbf{F}$ and $\xi_{A}=\mathcal{D}_{A}\mathbf{F}'$
have the same form, and their results $z^{k}$ and $z'^{k}$ describe
the same sequence $p^{k}$ on $M_{ST}$, if the Lorentz transformation
can be calculated exactly. \label{fig:Schematic_Cov} }
\end{figure}

Being both Lorentz covariant and symplectic, an algorithm is of significance
in two aspects. In the first place, the preservation of the symplectic
structure guarantees that the numerical solutions are good enough
to approximate the continuous solutions in arbitrary long time. Secondly,
the Lorentz covariance of algorithm makes the numerical results reference-independent,
which preserves the geometric nature of original systems. Figure \ref{fig:Schematic_Cov}
depicts the schematic diagram for the relation between a covariant
continuous system and the corresponding discrete systems generated
by the Lorentz covariant symplectic algorithm $A$. The 4-spacetime
is denoted by $M_{ST}$. The continuous evolution of the original
system forms a worldline, marked by $C_{wl}$, starting from the initial
condition $p^{0}$. The reference frames $\mathcal{O}$ and $\mathcal{O}'$
are two chosen Lorentz inertial frames. For a covariant continuous
system, the master equations $\mathbf{F}'$ in $\mathcal{O}'$ has
identical form as $\mathbf{F}$ in $\mathcal{O}$. According to the
Lorentz covariance, the solutions of $\mathbf{F}$ and $\mathbf{F}'$,
i.e., $z^{\tau}$ and $z'^{\tau}$, express the same worldline in
$M_{ST}$. Given $A$ is a covariant algorithm, the corresponding
discretized equations in $\mathcal{O}'$ is expressed as $\xi_{A}=\mathcal{D}_{A}\mathbf{F}'$,
and $\phi_{A}=\mathcal{D}_{A}\mathbf{F}$ denotes the discretized
equations in frame $\mathcal{O}$. The sequence determined by $\phi_{A}$
in $\mathcal{O}$ is denoted by $z^{k}$, and the sequence determined
by $\xi_{A}$ in $\mathcal{O}'$ in denoted by $z'^{k}$. According
to Eq.\,\ref{eq:CovAlgorithmDef}, we have the relation $\xi_{A}=\mathcal{T}_{L}\phi_{A}$
and thus $z'^{k}=\mathcal{T}_{L}z^{k}$ for each proper-time step
$k$. Consequently, as the analogy with the continuous case, the numerical
results of $A$ in different Lorentz frames provide different numerical
solutions $z^{k}$ and $z'^{k}$ but the same 4-worldpoint sequence
$p^{k}$ in $M_{ST}$. On the other hand, because $A$ is a symplectic
algorithm, the conservation of discrete symplectic structure ensures
$p^{k}$ locates adjacent to the exact solution of the original continuous
system $C_{wl}$ in $M_{ST}$, see the purple curve in Fig.\,\ref{fig:Schematic_Cov}.
We can conclude that the Lorentz covariant symplectic algorithms have
long-term numerical conservativeness, accuracy, and stability, which
are independent of the choice of reference frames.

\section{Construction of LCCSA\label{sec:Construction-CCSA}}

In this section, we introduce a convenient procedure for the construction
of LCCSA. The construction of an explicit LCCSA for relativistic dynamics
of charged particles is introduced step by step for demonstration.
This procedure can be generally applied for the construction of Lorentz
covariant symplectic algorithms for any other Lorentz covariant continuous
Hamiltonian systems. Since the Lorentz covariance should be preserved
during the discretization, the geometric properties in 4-spacetime
should be preserved. It is convenient to employ the Lorentz-covariant
forms of the continuous system to construct LCCSA.

Firstly, write explicitly down the covariant Hamiltonian equations
for charged particles in 4-spacetime. The covariant Hamiltonian describing
charged particle dynamics in electromagnetic fields is \cite{Goldenstein_ClassicalMech}
\begin{equation}
H=\frac{g^{\alpha\beta}\left(P_{\alpha}-A_{\alpha}\right)\left(P_{\beta}-A_{\beta}\right)}{2}\,,\label{eq:H4}
\end{equation}
where $X^{\alpha}$ is the 4-position vector, $P_{\alpha}$ is the
canonical momentum 1-form, and $A_{\alpha}$ denotes the 4-vector-potential
1-form. In Cartesian coordinate system, we have $X^{\alpha}=\left(t,\mathbf{x}\right)$,
$P_{\alpha}=\left(\gamma+\phi,-\mathbf{P}\right)$, $A_{\alpha}=\left(\phi,-\mathbf{A}\right)$,
and 
\[
g_{\alpha\beta}=g^{\alpha\beta}=\left(\begin{array}{cccc}
1 & 0 & 0 & 0\\
0 & -1 & 0 & 0\\
0 & 0 & -1 & 0\\
0 & 0 & 0 & -1
\end{array}\right)\,,
\]
where $\mathbf{P}$ is the canonical momentum, and $\phi$ and $\mathbf{A}$
are respectively the scalar and vector potentials of electromagnetic
fields. Before deriving the Hamiltonian equations, one should notice
that the evolution parameters should be Lorentz scalars, which is
vital to keep the Lorentz invariance of step-length after discretization.
As a direct consideration, we choose the proper time $\tau$ as the
evolution parameter. Correspondingly, according to the Hamiltonian
given in Eq.\,\ref{eq:H4}, we obtain the covariant Hamiltonian equations
$\mathbf{F}_{L4}$ \cite{Goldenstein_ClassicalMech,Jackson_electrodynamics},
\begin{equation}
\frac{\mathrm{d}P_{\alpha}}{\mathrm{d}\tau}=-\frac{\partial H}{\partial X^{\alpha}}=\left(P^{\beta}-A^{\beta}\right)\partial_{\alpha}A_{\beta}\,,\label{eq:Hamiltonian_P}
\end{equation}
\begin{equation}
\frac{\mathrm{d}X^{\alpha}}{\mathrm{d}\tau}=\frac{\partial H}{\partial P_{\alpha}}=P^{\alpha}-A^{\alpha}\,,\label{eq:Hamiltonian_X}
\end{equation}
where $\partial_{\alpha}=\partial/\partial X^{\alpha}=\left(\partial/\partial X^{0},\nabla\right)$,
$P^{\alpha}=g^{\alpha\beta}P_{\beta}$, and $A^{\alpha}=g^{\alpha\beta}A_{\beta}$.
It is readily to see that Eqs.\,\ref{eq:Hamiltonian_P} and \ref{eq:Hamiltonian_X}
are geometric equations and have reference invariant forms in all
Lorentz inertial frames. 

Secondly, discretize the Hamiltonian equations by using a canonical
symplectic method. To obtain an explicit scheme with high efficiency,
here we choose the Euler-symplectic method, which can be expressed
by \cite{Geometric_numerical_integration,CSPIC_2016}
\begin{equation}
P^{k+1}=P^{k}-h\frac{\partial H}{\partial X}\left(P^{k+1},X^{k}\right)\,,\label{eq:SEularP}
\end{equation}
\begin{equation}
X^{k+1}=X^{k}+h\frac{\partial H}{\partial P}\left(P^{k+1},X^{k}\right)\,,\label{eq:SEularX}
\end{equation}
where $h$ is the step-length. The Euler-symplectic method does not
break the geometric object or the form of continuous equations. Combining
Eqs.\,\ref{eq:Hamiltonian_P}-\ref{eq:SEularX}, we can obtain the
discrete equations of the LCCSA $\phi_{LCCSA}=\mathcal{D}_{LCCSA}\mathbf{F}_{L4}$
as 
\begin{equation}
P_{\alpha}^{k+1}=P_{\alpha}^{k}+\Delta\tau\left(P^{\beta,k+1}-A^{\beta,k}\right)\frac{\partial A_{\beta}^{k}}{\partial X^{\alpha}}\thinspace,\label{eq:CCSAP}
\end{equation}

\begin{equation}
X^{\alpha,k+1}=X^{\alpha,k}+\Delta\tau\left(P^{\alpha,k+1}-A^{\alpha,k}\right)\thinspace,\label{eq:CCSAX}
\end{equation}
where $\Delta\tau$ is the step-length of proper time. The difference
equations, Eqs.\,\ref{eq:CCSAP}-\ref{eq:CCSAX}, act as one-step
maps of geometric objects $\left(X^{\alpha,k},P_{\alpha}^{k}\right)\mapsto\left(X^{\alpha,k+1},P_{\alpha}^{k+1}\right)$.
As a property of geometric equations, Eqs.\,\ref{eq:CCSAP}-\ref{eq:CCSAX}
naturally inherit the reference-independence of Eqs.\,\ref{eq:Hamiltonian_P}-\ref{eq:Hamiltonian_X}.
The Lorentz covariance of the LCCSA can also be verified directly
through the definition Eq.\,\ref{eq:CovAlgorithmDef}. The Lorentz
transformation of $\phi_{LCCSA}$, $\phi_{LCCSA}'=\mathcal{T}_{L}\phi_{LCCSA}$,
can be given by left-multiplying the Lorentz matrix on both sides
of Eqs.\,\ref{eq:CCSAP} and \ref{eq:CCSAX}. Considering the linear
relations of all the terms in Eqs.\,\ref{eq:CCSAP}-\ref{eq:CCSAX},
it is obvious to see that $\phi_{LCCSA}'$ has the same form with
$\xi_{LCCSA}=\mathcal{D}_{LCCSA}\circ\mathcal{T}_{L}\mathbf{F}_{L4}$.
Therefore, LCCSA satisfies the definition of Lorentz covariant algorithms. 

During discretization, the Lorentz covariance cannot be inherited
without keeping geometric objects in 4-spacetime, even though the
4-dimentional covariant Hamiltonian equations are used. To explain
this, we provide a counter-example, a non-covariant algorithm (NCOVA)
of Eqs.\,\ref{eq:Hamiltonian_P}-\ref{eq:Hamiltonian_X}, namely,
$\phi_{NCOVA}$,
\begin{equation}
P_{\alpha}^{k+1}=P_{\alpha}^{k}+\Delta\tau\left(P^{\beta,k}-A^{\beta,k}\right)\partial_{\alpha}A_{\beta}^{k}\thinspace,\label{eq:NCOVAP}
\end{equation}
\begin{equation}
X^{0,k+1}=X^{0,k}+\Delta\tau\left(P^{0,k+1}-A^{0,k}\right)\thinspace,\label{eq:NCOVAX0}
\end{equation}
\begin{equation}
\mathbf{x}^{k+1}=\mathbf{x}^{k}+\Delta\tau\left(\mathbf{P}^{k}-\mathbf{A}^{k}\right)\thinspace,\label{eq:NCOVAX}
\end{equation}
where $X^{0}$, $P^{0}$, and $A^{0}$ denote the 0-components of
$X^{\alpha}$, $P^{\alpha}$, and $A^{\alpha}$, respectively. The
one-step map of $P_{\alpha}$ determined by Eq.\,\ref{eq:NCOVAP}
is the Euler method. In Eqs.\,\ref{eq:NCOVAX0}-\ref{eq:NCOVAX},
the 4-canonical-momentum for pushing $X^{\alpha}$ is treated in different
ways. When calculating $X^{0,k+1}$, $P^{0,k+1}$ is used. And $\mathbf{P}^{k}$
is used to calculate $\mathbf{x}^{k+1}$. The integrity of 4-dimentional
1-form $P_{\alpha}$ in Eq.\,\ref{eq:Hamiltonian_X} is thus broken,
which lead to different forms of Eqs.\,\ref{eq:NCOVAX0}-\ref{eq:NCOVAX}
after Lorentz transformations. The bad performance of this NCOVA under
Lorentz transformation is presented in numerical examples in Sec.\,\ref{sec:Drawbacks-of-Non-covariant},
which shows numerically that $\mathcal{T}_{L}\phi_{NCOVA}\neq\xi_{NCOVA}$,
where $\xi_{NCOVA}=D_{NCOVA}\circ\mathcal{T}_{L}\mathbf{F}_{L4}$.

\section{Numerical Experiments\label{sec:Drawbacks-of-Non-covariant}}

In this section, we analyze and test the performances of LCCSA through
several numerical experiments.

\subsection{The Lorentz covariance}

To test the Lorentz covariance of algorithms, the motion of an electron
is simulated in different Lorentz frames. The background magnetic
field is given by
\begin{equation}
\mathbf{B}=B_{0}\frac{R}{R_{0}}\mathbf{e}_{z}\,,\label{eq:CircleB}
\end{equation}
which has the vector potential
\begin{equation}
\mathbf{A}=B_{0}\frac{R^{2}}{3R_{0}}\mathbf{e}_{\theta}\,,\label{eq:CircleA}
\end{equation}
where $R=\sqrt{x^{2}+y^{2}}$, $\mathbf{e}_{z}$ and $\mathbf{e}_{\theta}$
are the unit vectors of cylindrical coordinates. The parameters of
field are set as $B_{0}=1\,\mathrm{T}$ and $R_{0}=\mathrm{m_{0}c}/\mathrm{e}B_{0}\approx1.69\times10^{-3}\,\mathrm{m}$.
We mark the lab reference frame as $\mathcal{O}$, where the initial
condition of the charged particle is set as $\mathrm{x}_{0}=\left(0,2R_{0},0\right)$
and $\mathbf{p}_{0}=\left(0,\mathrm{m_{0}c},0\right)$. We then find
another frame $\mathcal{O}'$ moves with velocity $\bm{\beta}_{cor}=\left(0.5,0,0\right)$
relative to $\mathcal{O}$. Initially, the local time of the $\mathcal{O}$
and $\mathcal{O}'$ are both set to be $0$, and the origin points
of $\mathcal{O}$ and $\mathcal{O}'$ coincide in 4-spacetime.

\begin{figure}
\includegraphics[scale=0.6]{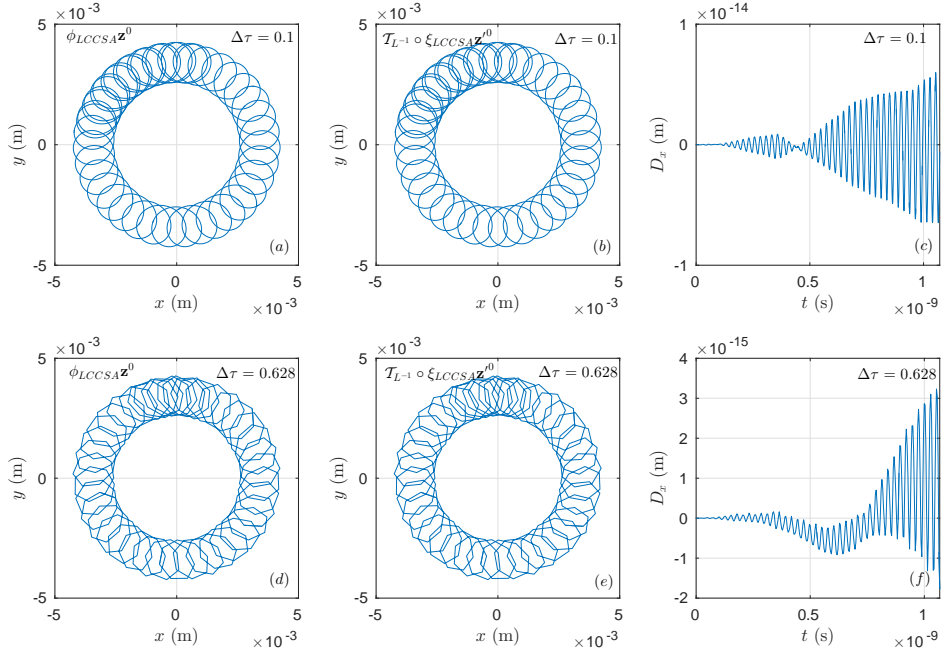}

\caption{The comparison of simulation results are given by LCCSA in different
Lorentz frames. Subfigures a, b and c are simulated with the step-length
$\Delta\tau=0.1$, while subfigures d, e and f are calculated by $\Delta\tau=0.628$.
The difference between the results calculated in two frames is in
the order of machine precision, which caused by the imprecision of
Lorentz transformation instead of the algorithm itself.\label{fig:CCSAcov}}
\end{figure}

In the case of LCCSA, we first apply $\xi_{LCCSA}$ in $\mathcal{O}'$.
Once obtained the numerical solution $\mathbf{z}'{}_{\xi}^{k}$ in
$\mathcal{O}'$, we transform it back to the frame $\mathcal{O}$
to get the result of $\mathbf{z}_{\xi}^{k}=\mathcal{T}_{L^{-1}}\circ\xi_{LCCSA}\mathbf{z}'^{0}$,
where $\mathbf{z}=\left(\mathbf{x},\mathbf{p}\right)$. On the other
hand, by using $\phi{}_{LCCSA}$, we can get discrete solution $\mathbf{z}_{\phi}^{k}=\phi_{CCSA}\mathbf{z}^{0}$
in $\mathcal{O}$ directly. The orbits of the electron in the x-y
plane are plotted in Fig.\,\ref{fig:CCSAcov}, and the difference
between the x-components of $\mathbf{z}_{\phi}^{k}$ and $\mathbf{z}_{\xi}^{k}$
is denoted by $D_{x}^{k}=x_{\phi}^{k}-x{}_{\xi}^{k}$. It can be observed
that the numerical difference comes from calculations in different
Lorentz frames is about $10^{-15}\,\mathrm{m}$, which is in the order
of machine precision. Meanwhile, $D_{x}^{k}$ is nearly independent
with the step-length, see Figs.\,\ref{fig:CCSAcov}c and Figs.\,\ref{fig:CCSAcov}f.
It is shown in Fig.\,\ref{fig:CCSAcov} that the difference equations
of LCCSA in $\mathcal{O}$ and $\mathcal{O}'$, namely, $\phi_{LCCSA}$
and $\xi_{LCCSA}$, can produce the same results if the numerical
error caused by the calculation of Lorentz transformation is neglected.
As a result, the stability, convergence, and consistency of LCCSA
are reference independent, which makes it safe to use LCCSA directly
in different frames.

\begin{figure}
\includegraphics[scale=0.6]{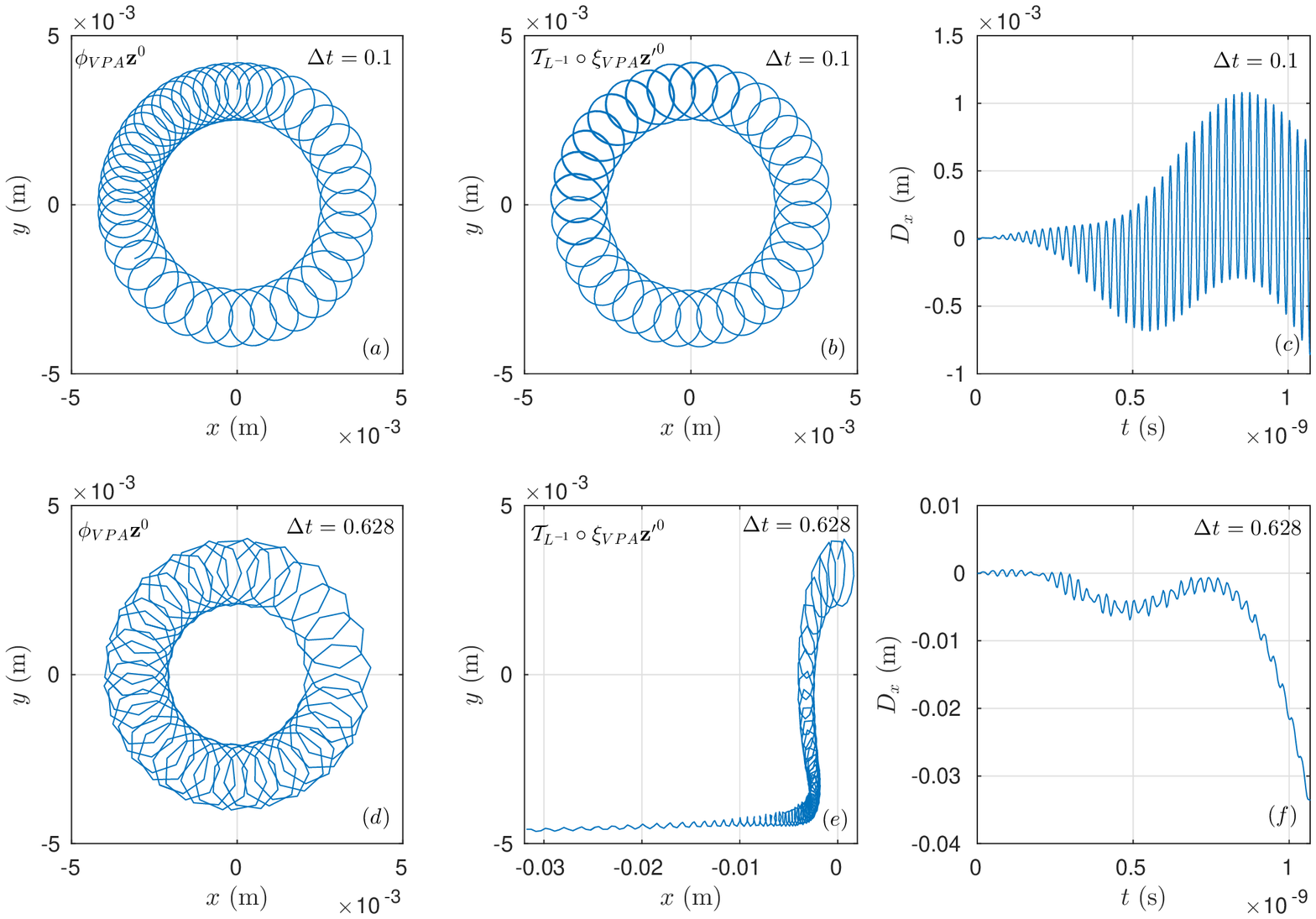}

\caption{The comparison of simulation results are given by VPA in different
Lorentz frames. Subfigures a, b and c are simulated with the step-length
$\Delta t=0.1$, while subfigures d, e and f are calculated with $\Delta t=0.628$.
The position difference between the numerical results in two frames
is comparable to $R_{0}$ with $\Delta t=0.1$. The VPA applied in
the frame $\mathcal{O}'$ turns out unstable with $\Delta t=0.628$.\label{fig:VPAcov}}
\end{figure}

\begin{figure}
\includegraphics[scale=0.65]{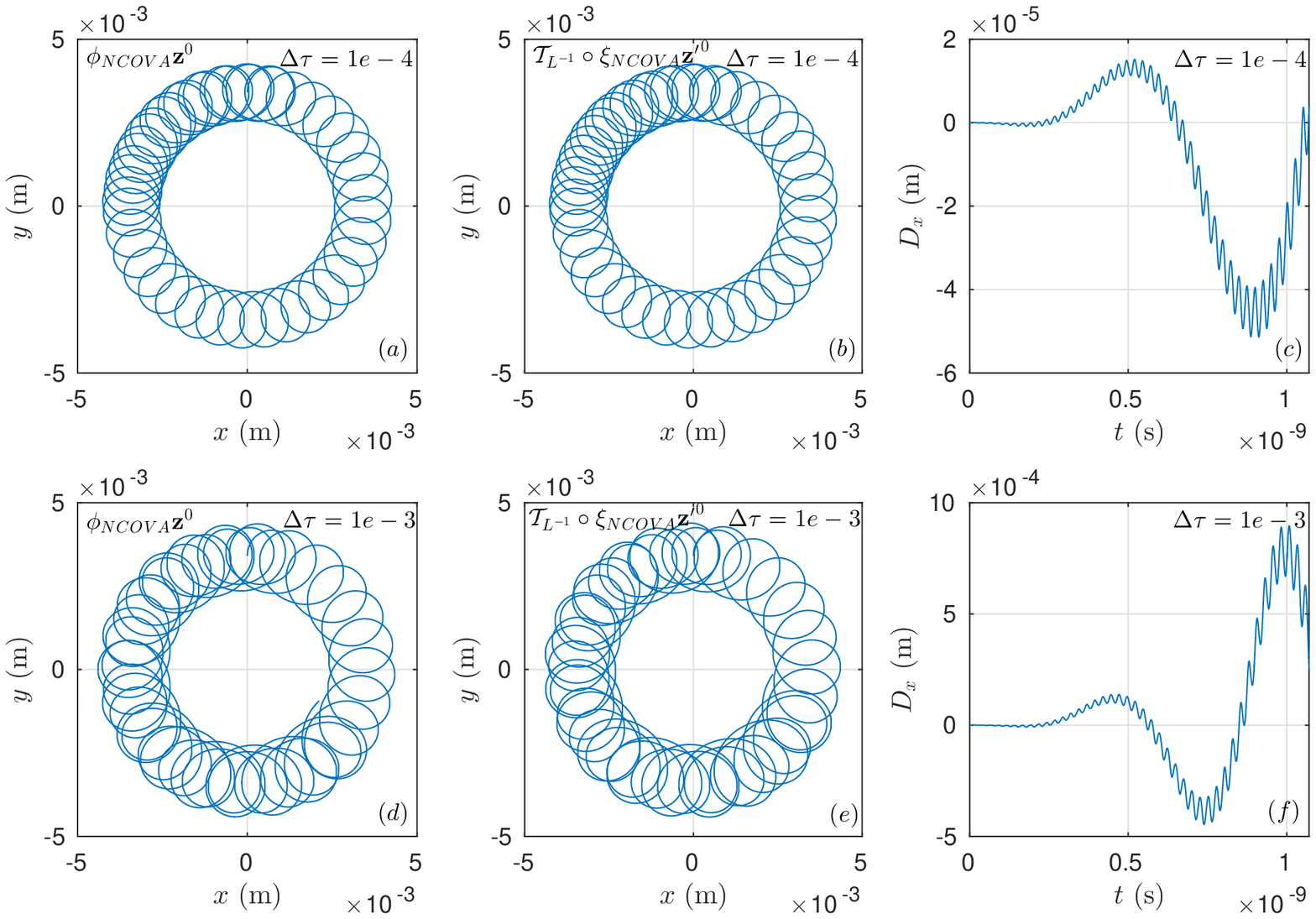}

\caption{The comparison of simulation results are given by NCOVA in different
Lorentz frames. Subfigures a, b and c are simulated with the step-length
$\Delta\tau=0.0001$, while subfigures d, e and f are calculated with
$\Delta\tau=0.001$. The NCOVA is a 1st-order non-covariant algorithm
and does not preserve symplecitc structure. The time step-length required
for stability is much smaller than LCCSA. The inconsistence between
the results calculated in different frames is comparable to $R_{0}$
though with small time steps.\label{fig:NCOVA}}
\end{figure}

For comparison, the relativistic VPA and NCOVA are also used to calculate
the same case. Because VPA is not a covariant algorithm as discussed
in Sec.\,\ref{sec:Lorentz-Covariant-Algorithms}, if we calculate
the dynamics of a charged particle in $\mathcal{O}'$ by use of $\mathbf{\xi}_{VPA}$,
its results $\mathbf{z}'{}_{\xi}^{k}$ cannot be simply transformed
back to the results $\mathbf{z}_{\phi}^{k}$ given by $\phi{}_{VPA}$
in $\mathcal{O}$, namely, $\mathcal{T}_{L^{-1}}\mathbf{z}'{}_{\xi}^{k}\neq\mathbf{z}_{\phi}^{k}$.
In other words, if observing in $\mathcal{O}$, the VPA carried out
in different reference frames $\mathcal{T}_{L^{-1}}\circ\xi_{VPA}\mathbf{z}'^{0}$
and $\phi_{VPA}\mathbf{z}^{0}$ are actually two different algorithms
with different properties and outputs. With the same field configuration
and initial conditions, the results from $\xi_{VPA}$ and $\phi_{VPA}$
are shown in Fig.\,\ref{fig:VPAcov}. If $\Delta t=0.1$, see Fig.\,\ref{fig:VPAcov}c,
the position difference of orbits in Fig.\,\ref{fig:VPAcov}a and
b is in the order of $R_{0}\sim10^{-3}\,\mathrm{m}$. If $\Delta t=0.628$,
see Fig.\,\ref{fig:VPAcov}e, $\mathcal{T}_{L^{-1}}\circ\xi_{VPA}\mathbf{z}'^{0}$
becomes unstable and gives wrong numerical results. Similarly, the
non-covariant property of NCOVA is shown in Fig.\,\ref{fig:NCOVA}.
When applied in different frames, NCOVA also becomes different algorithms
and hence has different performances, see numerical results $\phi_{NCOVA}\mathbf{z}^{0}$
and $\mathcal{T}_{L^{-1}}\circ\xi_{NCOVA}\mathbf{z}'^{0}$ in Fig.\,\ref{fig:NCOVA}a,
b, d, and e. The $D_{x}$ is also comparable to the value of $R_{0}$
and dependent with the step-length, see Fig.\,\ref{fig:NCOVA}e,
f. According to Figs.\,\ref{fig:VPAcov} and \ref{fig:NCOVA}, the
non-covariant problem results from the non-covariant algorithms in
different Lorentz frame are well exhibited.

\subsection{The secular stability}

All LCCSAs possess good long-term properties belonging to standard
symplectic algorithms. The covariant Hamiltonian in Eq.\,\ref{eq:H4},
known as the mass-shell, is a constant of motion. Through conserving
the symplectic structure, LCCSA can restrict the global error of the
mass-shell under a small value \cite{Geometric_numerical_integration}.
For comparison, we develop a Lorentz covariant but non-symplectic
algorithm, i.e., a fourth-order Runge-Kutta method (RK4), to solve
the 4-dimentional covariant Lorentz equations,
\begin{equation}
\frac{\mathrm{d}X^{\alpha}}{\mathrm{d}\tau}=U^{\alpha}\,,\label{eq:Lorentz4D_X}
\end{equation}
\begin{equation}
\frac{\mathrm{d}p^{\alpha}}{\mathrm{d}\tau}=F^{\alpha\beta}U_{\beta}\,,\label{eq:Lorentz4D_P}
\end{equation}
where $p^{\alpha}$ is the 4-mechanical-momentum, $U^{\alpha}$ is
the 4-velocity, and $F^{\alpha\beta}$ is the electromagnetic tensor
\cite{Jackson_electrodynamics}. We can see that RK4 is a Lorentz
covariant algorithm because its discretization does not break the
geometric structure of Eqs.\,\ref{eq:Lorentz4D_X} and \ref{eq:Lorentz4D_P}.

\begin{figure}
\includegraphics{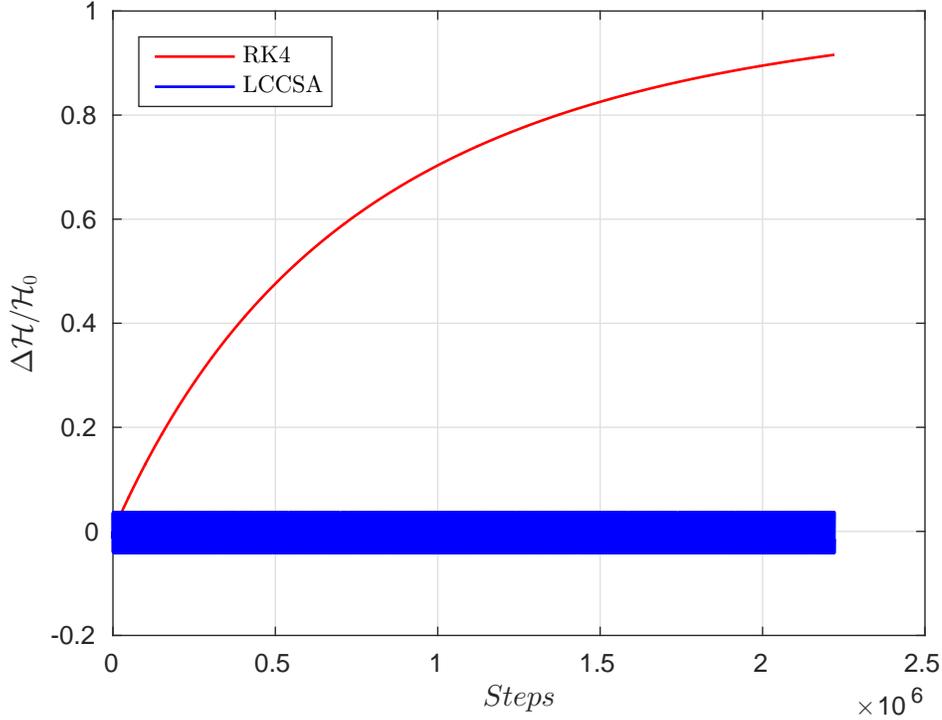}

\caption{The evolutions of relative errors of mass-shell by RK4 and LCCSA.
The step-length is $\Delta\tau=0.1$. The error of mass-shell given
by RK4 becomes comparable to $\mathcal{H}_{0}$ after $2\times10^{6}$
steps, while the relative error is limited under a small value in
the case of LCCSA. \label{fig:Herror}}
\end{figure}

Figure \ref{fig:Herror} compares the evolutions of relative numerical
error of mass-shell calculated by RK4 and the LCCSA. The electromagnetic
field and initial conditions are set the same as in Fig.\,\ref{fig:CCSAcov},
and the step-length is set to be $\Delta\tau=0.1$. After $2\times10^{6}$
proper-time steps, the relative mass-shell error of RK4 accumulates
to a significant value, which results in unreliable numerical results.
However, the relative error of LCCSA keeps bounded in a small region
due to its symplectic nature. According to this numerical experiment,
Lorentz covariant algorithms without secular conservativeness suffer
from coherent accumulation of numerical errors, which implies the
necessity to combine the Lorentz covariance and the structure-preserving
methods.

\subsection{The energy-based adaptive time step}

Through the discretization of the proper time $\tau$, LCCSA also
possesses built-in energy-based adaptive-time-step property. The discrete
relation between $\Delta\tau$ and $\Delta t$ can be reflected by
the 0th component of Eq.\,\ref{eq:CCSAX} as

\begin{equation}
\Delta t=t^{k+1}-t^{k}=\Delta\tau\left(P_{0}^{k+1}-\phi^{k}\right)\,,\label{eq:dtanddtau}
\end{equation}
where $P_{0}$ is the 0th component of canonical momentum, and $\phi$
is the electric potential. Considering that the expression in the
bracket of Eq.\,\ref{eq:dtanddtau} can be rewritten as $P_{0}^{k+1}-\phi^{k}=\gamma^{k+1}+\phi^{k+1}-\phi^{k}$
and $\Delta\phi=\phi^{k+1}-\phi^{k}$ generally is a small value,
the time step $\Delta t$ is approximately proportional to $\gamma^{k+1}$.
Equation \ref{eq:dtanddtau} is actually a discrete version of the
relation $\mathrm{d}t=\gamma\mathrm{d}\tau$. For constant $\Delta\tau$,
the time step $\Delta t$ can be self-adapted according to the energy
of particles. When a charged particle moves in an extern magnetic
field, the gyro-period $T_{ce}=2\pi\gamma\mathrm{m}_{0}/\mathrm{e}B$
determines the smallest time-scale of the particle dynamics. In simulations,
to resolved the dynamical behaviors smaller the time-scale of gyro-period,
the time step should be restricted smaller than $T_{ce}$. When considering
the efficiency of computation, too small time step brings heavy computation
consuming. One should choose a suitable $\Delta t$ to balance the
accuracy and the efficiency. Because $T_{ce}$ is proportional to
$\gamma$, for algorithms with fixed time step $\Delta t$, time steps
lie in one gyro-period grows as the increase of $\gamma$, which cause
the waste of calculation resources in problems with increasing $\gamma$.
On the other hand, if the particle loses energy quickly in some processes,
$T_{ce}$ may drop to smaller than $\Delta t$, which results in numerical
instabilities for algorithms. However, the time-step problems can
be avoided easily by using the LCCSA. 

\begin{figure}
\includegraphics[scale=0.8]{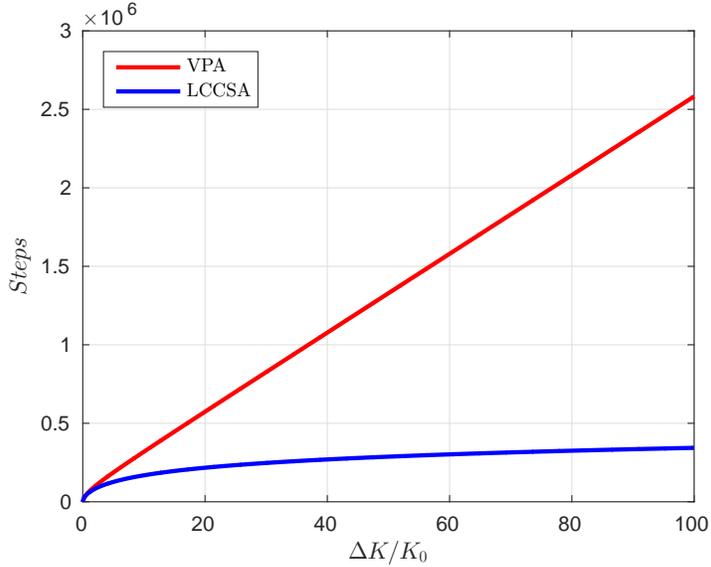}

\caption{The numbers of simulation steps in terms of the relative increase
of energy required by VPA and LCCSA to simulate the same acceleration
process of an electron.\label{fig:Aclr}}
\end{figure}

\begin{figure}
\includegraphics[scale=0.8]{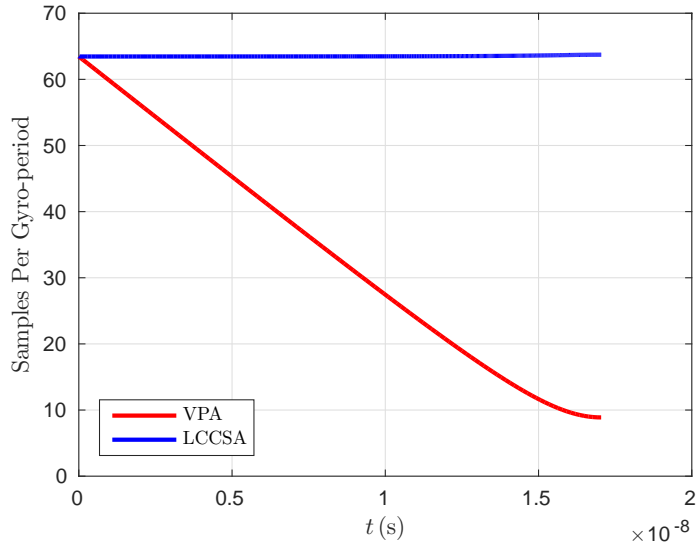}

\caption{Numbers of time steps in one gyro-period when employing VPA and LCCSA
to simulate the same decelerate process of an electron. As the decrease
of the energy, the number of time-samplings in one gyro-period for
LCCSA keeps unchanged, while the number of time-samplings for VPA
decreases.\label{fig:Brake}}
\end{figure}

To show the advantages of the energy-based adaptive time steps, the
acceleration and braking process of an electron is simulated in a
uniform magnetic field. We compare the performance of LCCSA with VPA
which has a fixed time step \cite{Ruili_VPA_2015}. Both the electric
and magnetic fields have only z-component, namely, $\mathbf{B}=B_{0}\mathbf{e}_{z}$
and $\mathbf{E}=E_{0}\mathbf{e}_{z}$. In the acceleration process,
the particle is released at $x=1.8\,\mathrm{m}$, $y=z=0$, the magnetic
field is set as $B_{0}=2\,\mathrm{T}$, and the electric field is
$E_{0}=10000\,\mathrm{V/m}$. The initial momentum of the electron
is given by $\mathbf{p}_{0}=\left(0,1\,\mathrm{m_{0}c},0.1\,\mathrm{m_{0}c}\right)$.
Figure \ref{fig:Aclr} shows the number of steps iterated by VPA and
LCCSA in terms of different relative increments of kinetic energy.
As the increase of the energy, the slope of red curve keeps unchanged,
while the slope of blue curve decreases significantly, see Fig.\,\ref{fig:Aclr}.
Therefore, to reach the same energy, the computation efficiency of
LCCSA is much better than VPA. In the case of braking process, the
initial position and the magnetic field are the same as before, the
electric field is set as $E_{0}=1\,\mathrm{MV/m}$, and the initial
momentum is given by $\mathbf{p}_{0}=\left(0,\ 1\,\mathrm{m_{0}c},\ -10\,\mathrm{m_{0}c}\right)$.
Figure \ref{fig:Brake} depicts the number of time samplings during
each gyro-period. The sampling number of VPA in one gyro-period decreases
as the decrease of energy due to the fixed time step, while the time
sampling number of LCCSA keeps unchanged. In this case, through adjusting
the time step automatically, LCCSA can provide higher accuracy than
VPA and avoid numerical instabilities in the simulation of energy
decrease processes. 

\section{Conclusions\label{sec:Conclusions}}

In this paper, we provide the definition of Lorentz covariant algorithms
and introduce Lorentz covariant symplectic algorithms in detail. Lorentz
covariant algorithms can generate discretized equations, which inherits
the Lorentz covariant nature of original continuous systems. Symplectic
algorithms without Lorentz covariance only performs well in one specific
inertial frame. While covariant symplectic algorithms are reference-independent
and possess long-term conservativeness, which make it convenient and
safe to employ the same algorithm in any Lorentz frame. Because of
the essentiality of Lorentz covariance, the Lorentz covariant symplectic
algorithms have wide applications.

On the other hand, because the time-variable becomes a component of
coordinate for 4-spacetime in the construction of LCCSA, the time-dependent
Hamiltonian system is no longer a problem for the construction of
required symplectic algorithms. Taking the proper time $\tau$ as
the dynamical parameter, all time-dependent Hamiltonian system becomes
proper-time-independent. The explicit symplectic algorithm, like the
LCCSA in Eqs.\,\ref{eq:CCSAP}-\ref{eq:CCSAX}, for time dependent
systems can be easily constructed. According to the idea and procedure
in this paper, many other Lorentz covariant symplectic algorithms
as well as other kinds of Lorentz covariant structure-preserving algorithms
can be readily constructed. In the future work, we will further investigate
the Lorentz covariant structure-preserving algorithms and apply the
LCCSAs to study key physical problems.
\begin{acknowledgments}
This research is supported by National Magnetic Confinement Fusion
Energy Research Project (2015GB111003, 2014GB124005), National Natural
Science Foundation of China (NSFC-11575185, 11575186, 11305171), JSPS-NRF-NSFC
A3 Foresight Program (NSFC-11261140328), Key Research Program of Frontier
Sciences CAS (QYZDB-SSW-SYS004), and the GeoAlgorithmic Plasma Simulator
(GAPS) Project. 
\end{acknowledgments}

\bibliographystyle{apsrev}
\bibliography{Refs_RunawayElectrons}

\end{document}